\documentclass[prl,aps,twocolumn]{revtex4}
\usepackage{graphicx}
\usepackage{amsmath}

\def\ba{\begin{eqnarray}}
\def\ea{\end{eqnarray}}
\def\bse{\begin{subequations}\begin{align}}
\def\ese{\end{align}\end{subequations}}
\newcommand{\mc}[1]{\mathcal{#1}}
\newcommand{\hmc}[1]{\hat{\mathcal{#1}}}

\newcommand{\bra}[1]{\langle#1|}
\newcommand{\ket}[1]{|#1\rangle}

\begin{document}
\title{Nonlinear optics with stationary pulses of light}
\author{A. Andr\'{e}, M. Bajcsy, A. S. Zibrov, and M. D. Lukin}
\address{Physics Department, Harvard University, Cambridge,  MA 02138}
\date{\today}
\begin{abstract}
We show that the recently demonstrated technique for generating 
stationary pulses of light [Nature {\bf 426}, 638 (2003)] can be extended to localize optical pulses in all three spatial dimensions in a resonant atomic medium. This method can be used to dramatically enhance the nonlinear interaction between weak optical pulses.
In particular, we show that an efficient Kerr-like interaction between
two pulses can be implemented as a sequence of several purely linear optical processes. The resulting process may enable coherent 
interactions between single photon pulses. 
\end{abstract}
\maketitle


Techniques that could facilitate controlled  nonlinear interactions between few-photon light pulses are now actively explored \cite{lukin-nat}. Although research into fundamental limits of nonlinear optics 
has been carried out over the last three decades, there is renewed   
interest in these problems in part due to e.g. potential applications in quantum 
information science \cite{qinfo}. 
In general, such interactions between few-photon pulses are difficult to achieve, as they require
a combination of large nonlinearity, low photon loss and tight
confinement of  the light beams \cite{boyd}. In addition, long atom-photon interaction times are required. Simultaneous implementation of all of these  requirements is  by now only  feasible in the context of cavity QED \cite{cavqed}.

In this Letter we describe a novel method for achieving nonlinear interaction between weak light pulses. Our method is based on a recently demonstrated technique  \cite{axel,micho} in which light propagating in  a medium of Rb atoms was converted  into an excitation with localized, stationary electromagnetic energy, which could be held and released  after a controllable interval.  This is 
achieved by using Electromagnetically Induced Transparency (EIT) \cite{harris-phystoday} to 
coherently control the pulse propagation. We show here
that this method can be extended to confine stationary pulses in all three spatial dimensions.
This, in turn, can be used to strongly enhance
the nonlinear interaction between weak pulses of light. Specifically we
demonstrate  that an efficient Kerr-like interaction between
two  pulses can be implemented as a sequence of linear optical processes and atomic state manipulations. Coherent, controlled  nonlinear processes at optical energies corresponding to a single light  quanta  appear feasible.

Before proceeding, we note that the present work is closely 
related to recent studies on the resonant enhancement of 
nonlinear optical phenomena via EIT 
\cite{atac-nlo,harris-yoshi,kurizki,masalas}. 
The essence of these studies
is to utilize steep atomic dispersion associated with narrow EIT resonances. 
In such a system, a small AC Stark shift 
associated with  a weak off-resonant pulse of signal light, produces 
a large change in refractive index for a resonant probe pulse.
In order to fully take 
advantage of this process, long interaction times between signal and
probe pulses must be ensured. Although the latter  can be achieved by reducing the group velocities of two interacting  pulses by equal amounts \cite{lukin-ima}, 
in practice this results only in a modest increase of the nonlinear optical efficiency since reduction of the group velocity is accompanied
by a corresponding decrease of the light energy in the propagating pulse. Moreover, such nonlinear interaction is accompanied 
by pulse distortion,  which poses a fundamental limit to nonlinear interactions. 
In contrast, the technique presented here allows for  long interaction times (associated with stationary light pulses) without proportional reduction of the photon energy. Furthermore,  our technique enables light localization  in all three spatial dimensions in the presence of atoms,  and allows to entirely avoid competing effects such as pulse distortion. 


The ideas discussed in Refs. \cite{micho} allow one to localize and hold 
a pulse of light within a stationary envelope along the propagation direction. In practice, focused pulses will undergo diffraction in the transverse directions. In order to prevent diffraction, 
it is necessary to confine the signal beam in transverse directions. 
This can be achieved, e.g. by using a hollow core photonic crystal fiber filled with 
an active medium of resonant atoms \cite{pbg}.  Such an approach is particularly attractive in that it
ensures a single-mode beam quality for interacting beams. The 
unwanted interactions of atoms with fiber walls can be avoided
by using  atom guiding techniques \cite{atomguide}. 
Alternatively, the signal pulse guiding can be accomplished by using focused control beams. 
Shaped control beams can be used to create a transverse variation of the index of refraction,
enabling waveguiding and confinement of light pulses to small transverse dimensions.

To be specific we consider a medium of length $L$ consisting of an ensemble of $N$ three-level atoms in the $\Lambda$ configuration, with two metastable lower states, as shown in Fig.~1a. The ground states $\ket{g},\ket{s}$ are coupled to the excited state $\ket{e}$ via a control field applied on resonance with the $\ket{s}\rightarrow\ket{e}$ transition and a weak quantized signal field close to resonance with the $\ket{g}\rightarrow\ket{e}$ transition.
The control field consists of two counter-propagating fields with spatially and temporally varying Rabi frequencies $\Omega_\pm(\vec{r},t)$, so that the control field Rabi frequency is $\Omega(\vec{r},t)=\Omega_+(\vec{r},t)e^{ik_cz}+\Omega_-(\vec{r},t)e^{-ik_cz}$, where $k_c=n_c\omega_{es}/c$, $n_c$ being the background index of refraction at the frequency $\omega_{es}$.
The corresponding signal fields have slowly varying envelopes $\hat{E}_\pm(\vec{r},t)$, so that the signal field is
\ba
\hat{E}_S^{(+)}(\vec{r},t) &=& 
\left(\frac{\hbar\omega_0}{2\epsilon_0V}\right)^{1/2}\left[\hat{E}_+(\vec{r},t)e^{ik_sz}
\right.\nonumber\\
&+& \left.
\hat{E}_-(\vec{r},t)e^{-ik_sz}\right]e^{-i\omega_{eg}t},
\ea 
where $k_s= n_s \omega_{eg}/c$, $V$ is the quantization volume, and $n_s$ is the background refractive index at the frequency $\omega_{eg}$ due to off-resonant atomic levels. Note that in practice  $n_s$ can be tuned, e.g. by changing the light polarization. 

We describe the atomic properties with slowly varying collective operators \cite{fleischhauer00} 
$\hat{\sigma}_{\mu\nu}(\vec{r},t)=
\frac{1}{N_{\vec{r}}}\sum_{j=1}^{N_{\vec{r}}}\ket{\mu}_j\bra{\nu}e^{-i\omega_{\mu\nu}t}$
where $\omega_{\mu\nu}=(E_{\mu}-E_{\nu})/\hbar$, and where $N_{\vec{r}}$ is the number of atoms in a small but macroscopic volume around position $\vec{r}$.
We define the polarization operator to be $\hat{P}(\vec{r},t)=\sqrt{N}\hat{\sigma}_{ge}(\vec{r},t)$, and the spin flip operator $\hat{S}(\vec{r},t)=\sqrt{N}\hat{\sigma}_{gs}(\vec{r},t)$.
In the present situation of weak signal fields and strong control fields, most atoms are in state $\ket{g}$, with a few spin-flipped atoms in $\ket{s}$, so that the polarization and spin flip operators obey bosonic commutation relations \cite{fleischhauer00}.
Associated with the forward/backward
propagating fields are slowly varying polarization envelopes
$\hat{P}_\pm(\vec{r},t)$, so that the total polarization is
$\hat{P}(\vec{r},t)=\hat{P}_+(\vec{r},t)e^{ik_sz}+\hat{P}_-(\vec{r},t)
e^{-ik_sz}$.

\begin{figure}
\includegraphics[width=8.5cm]{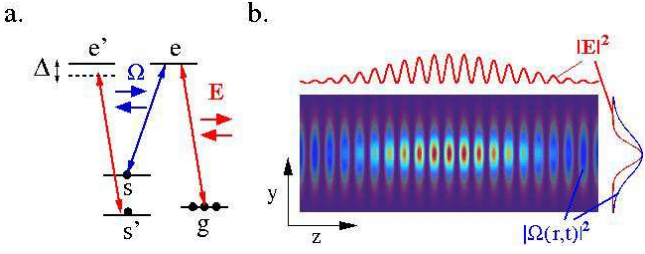}
\caption{
a. Three-level atoms in $\Lambda$-configuration, with auxiliary levels $\ket{s'},\ket{e'}$. 
Forward and backward propagating control fields with Rabi frequencies $\Omega_\pm$, and weak signal field $\mc{E}_\pm$.
b. Three-dimensional confinement and waveguiding of signal light due to transverse intensity profile and longitudinal modulation of control field.}
\end{figure} 

Letting the wavevector mismatch between co-propagating signal and control fields be $\Delta K=k_s-k_c=(n_s\omega_{eg} -n_c\omega_{es})/c$, and defining $\hmc{E}_\pm=\hat{E}_{\pm}e^{\pm i\Delta Kz}$ and $\hmc{P}_\pm=\hat{P}_\pm e^{\pm i\Delta Kz}$, the Heisenberg equations of motion for the slowly varying operators $\hmc{E}_\pm(\vec{r},t)$ \cite{fleischhauer00} can be written as (in the paraxial approximation)
\ba
\left(\frac{\partial}{\partial t}\pm c\frac{\partial}{\partial z}-i\frac{c\nabla_T^2}{2k_s}\right)\hmc{E}_\pm
&=& i\Delta Kc\hmc{E}_\pm+ig\sqrt{N}\hmc{P}_\pm,
\ea
where $\nabla_T^2=\nabla^2-\frac{d^2}{dz^2}$ is the transverse Laplacian, and where $g=\wp\left(\frac{\omega_0}{2\hbar\epsilon_0V}\right)^{1/2}$ is the atom-field coupling constant, $\wp$ being the dipole matrix element for the $\ket{g}-\ket{e}$ transition, and $V$ the quantization volume.

Following   \cite{fleischhauer00,axel} we introduce  two components
$\hat{\Psi}_\pm$ of a coupled excitation of light and an atomic spin wave ("dark-state polariton")
corresponding to forward and backward signal fields respectively. 
In the experimentally relevant case of  small group 
velocities \cite{hau99,liu01} the polariton components are represented by 
$\hat{\Psi}_{\pm} = g\sqrt{N}\hmc{E}_{\pm}/\Omega_{\pm}$. 
In the adiabatic limit of slowly varying pulses, disregarding the slow decay of ground state coherence, 
and Fourier transforming ($\partial_t\rightarrow -i\omega$), we find
\begin{subequations}
\label{eq:3dstateqs}
\begin{align}
\left(c\frac{\partial}{\partial z}-i\frac{c\nabla_T^2}{2k_0}\right)\hat{\Psi}_+
&= 
i\Delta{K}c\hat{\Psi}_+
\nonumber\\
+i\eta\omega\left(\alpha_+\hat{\Psi}_++\alpha_-\hat{\Psi}_-\right)
&-\alpha_-\xi\left(\hat{\Psi}_+-\hat{\Psi}_-\right)+\hat{F}_+(\vec{r},\omega)
\\
\left(-c\frac{\partial}{\partial z}-i\frac{c\nabla_T^2}{2k_0}\right)\hat{\Psi}_-
&= 
i\Delta{K}c\hat{\Psi}_-
\nonumber\\
+i\eta\omega\left(\alpha_+\hat{\Psi}_++\alpha_-\hat{\Psi}_-\right)
&+\alpha_+\xi\left(\hat{\Psi}_+-\hat{\Psi}_-\right)+\hat{F}_-(\vec{r},\omega)
\end{align}\end{subequations}
where $k_0=\omega_{eg}/c$, $\eta=\frac{g^2N}{|\Omega_+|^2+|\Omega_-|^2}$, $\alpha_\pm=\frac{|\Omega_\pm|^2}{|\Omega_+|^2+|\Omega_-|^2}$, and $\xi=\frac{g^2N}{\gamma}$.
We have also assumed that $k_0 a\gg1$, where $a$ is the typical transverse size of the control beams. 
These equations describe two slow waves that are coupled
due to periodic modulation of atomic absorption  and group velocity.
The terms containing $\xi$ on the right hand side of Eqns. \eqref{eq:3dstateqs} are proportional to the absorption coefficient $\xi$ near resonant line center.
$\hat{F}_\pm(\vec{r},\omega)$ are noise forces associated with dissipation.
When $\xi$ is large these terms
give rise to the pulse matching phenomenon  \cite{harris-phystoday}: whenever one of the fields is created the other will adjust itself within a short propagation distance to match its amplitude such that $\hat{\Psi}_+-\hat{\Psi}_-\rightarrow 0$ \cite{micho}.

In order to achieve transverse confinement of light pulses, we take into account the transverse dependence of the control field intensity and the resulting variation of the index of refraction. 
For a focused control beam, the intensity decreases with distance from the optical axis, so that for negative two-photon detuning (as is necessary for phasematching), the index of refraction decreases with distance from the optical axis. This leads to waveguiding of the signal light.
 The combination of waveguiding with strong coupling of forward and backward propagating modes, permits the complete three-dimensional confinement of light pulses in the medium.

We assume a transverse spatial variation of the control field intensity, e.g. for a weakly focused gaussian beam, $|\Omega_\pm(r)|^2=|\Omega_\pm(0)|^2e^{-(r/a)^2}$. 
Expanding for $r\ll a$, we 
have $\eta(r)=\eta_0[1+(r/a)^2+\cdots]$.
We consider trial solutions of (\ref{eq:3dstateqs}) of the from 
$\Psi_+=A_+e^{i\beta z-(r/R)^2}$ and 
$\Psi_-=A_-e^{-i\beta z-(r/R)^2}$ where $\beta$, $R$ and 
$A_+/A_-$  are determined  by  requiring that 
the coefficients of different powers of $r$ vanish independently. 
With these requirements  we find the two 
solutions $R=\infty$ and
\ba
R &=& \left(-\frac{2a^2c}{k_0 \eta\omega}\right)^{1/4}.
\ea 
The eigenvector of the finite solution has $A_+/A_-=1$,  which corresponds to the stationary pulses \cite{micho} for which $\Psi_+-\Psi_-\rightarrow 0$. 
We also find the dispersion relation 
\ba
\eta\omega =
\left(\frac{2c}{k_0 R(\omega)^2}-\Delta{K}c\right)
-i\frac{(c\beta)^2}{\xi}+(\alpha_+-\alpha_-)c\beta.
\ea
In the time-domain this corresponds to propagation at a group velocity $v_g=c\frac{\alpha_+-\alpha_-}{\eta}$, that can be controlled by adjusting the intensities $I_\pm(t)\propto|\Omega_\pm(t)|^2$ of the counter-propagating control fields. Due to the imaginary term, there is also
a slow spreading of the stationary pulse at a rate $\delta l/l \sim \sqrt{c^2t/(\eta\xi l^2)}$, which determines the maximal trapping time of the stationary excitation.  

We are interested in a simultaneous solution of Eqs.(\ref{eq:3dstateqs}) when $\alpha_+ \approx \alpha_-$ and the optical depth $\xi$
is large. This yields the radius of the stationary pulse 
$R = a2^{1/4}\left[\sqrt{1+\Delta K k_0 a^2}-1\right]^{-1/2}$, which 
under conditions of strong confinements $\Delta K k_0 a^2 \gg 1$ results in 
\begin{equation}
R=a[2/(\Delta K k_0 a^2)]^{1/4}. 
\end{equation}

Hence, in an optically dense medium ($\xi L/c\gg 1$) a stationary excitation confined in all three dimensions can be controllably created.
To be specific, for atomic Rb ($\lambda = 0.8 \mu$m) at density $n=10^{14}{\rm cm}^{-3}$, we take
the background refractive index due to off-resonant levels to be $n-1=1.2\cdot10^{-2}$ \cite{sasha}.
For a Gaussian control beam with waist $a \sim100{\rm \mu m}$ at the center of the atomic cell ( the corresponding Rayleigh range is $z_0=3.9{\rm cm}$), we find that  the guided mode radius is $R=13{\rm \mu m}$ (for which the Rayleigh range is $z_0=0.06{\rm cm}$),
so that the diffraction-free range is extended by  a factor of $60$. 
When the control beam is chosen in the form 
 of a  non-diffracting  Bessel beam \cite{durnin}, 
 waveguiding over  much longer propagation distances is possible. For example, with a  beam in which the radius of the first lobe of the Bessel function is  $a=20{\rm \mu m}$, the guided mode radius is $R\sim 5.7 {\rm \mu m}$. In practice, this allows confinement over tens of cm, whereas in free space the corresponding Rayleigh range would only be $0.01{\rm cm}$.
Reducing the control beam radius until $R\sim a$, gives an estimate of the smallest guided mode achievable, which for an index of $n-1=1.2\cdot10^{-2}$ is $R_{min}=1.6{\rm \mu m}$.
 

We next turn to the nonlinear optical interaction between two weak pulses of light. A notable feature of the step by step process described below, is that it consists of a sequence of purely linear optical interactions and atomic state manipulations, leading to an effective optical nonlinear interaction.
As shown in the timing diagram Fig.~2a, a light pulse (signal pulse) travelling through the atomic medium is initially stored in the $\ket{g}\bra{s}$ coherence.
Next,
a Raman or microwave $\pi$ pulse (RA) transfers the population from $\ket{s}$ to $\ket{s'}$ (see Fig.~2a), thereby transferring the stored excitation to the spin excitation $\hmc{S'}=\sqrt{N}\hat{\sigma}_{gs'}$. 
The signal pulse is stored as a spin wave $\sigma_{gs'}(\vec{r},t)=U_s(\vec{\rho})\hmc{S}'(z,t)/\sqrt{N}$, where $U_s(\vec{\rho})$ describes the transverse spin wave mode.

A second light pulse (probe pulse) is then sent through the medium and stored in the coherence $\hmc{S}$.
The spin excitation associated with the probe pulse is then converted into a stationary excitation. The latter is then moved through the stored spin excitation associated with the signal pulse. 
Probe light in the modes 
$\hmc{E}_\pm$ interacts dispersively with atoms in level $\ket{s'}$ (see Fig.~1a), thereby acquiring a phase shift proportional to the number of excitations $\hmc{S'}^\dagger\hmc{S'}$, 
leading to an effective Kerr-type nonlinearity.

We focus on the situation where the transverse spatial size of the signal spin wave $\hmc{S}'$ is much smaller than the probe transverse size.
Under these assumptions and
writing the guided mode transverse dependence as $\hat{\Psi}_\pm(r,z,t)=e^{-(r/R)^2}\hat{\psi}_\pm(z,t)$, we find, in terms of the polariton components $\hat{\psi}_\pm$,
\begin{subequations}\begin{align}
(\partial_t+c\partial_z)\hat{\psi}_+ &=
-\eta\partial_t
(\alpha_+\hat{\psi}_++\alpha_-\hat{\psi}_-)-
\alpha_-\xi(\hat{\psi}_+-\hat{\psi}_-)
\nonumber\\
&+ i\beta\hmc{S'}^\dagger\hmc{S'}\hat{\psi}_+
+\hat{F}_+(z,t)
\\
(\partial_t-c\partial_z)\hat{\psi}_- &= 
-\eta\partial_t(\alpha_+\hat{\psi}_++\alpha_-\hat{\psi}_-)+
\alpha_+\xi(\hat{\psi}_+-\hat{\psi}_-)
\nonumber\\
&+i\beta\hmc{S'}^\dagger\hmc{S'}\hat{\psi}_-
+\hat{F}_-(z,t)
\end{align}\end{subequations}
where $\beta=\frac{\tilde{g}^2}{\Delta}(1+i\gamma/\Delta)$,
with $\tilde{g}=g\frac{A}{\pi R^2}$,
$A$ is the quantization area, and $\Delta$ is the 
detuning of the fields $\hmc{E}_\pm$ from the optical transition 
$\ket{s'}\rightarrow\ket{e'}$ (see Fig.~2a).
The interaction of the localized excitation $\hmc{S}'$ with the guided modes $\hat{\psi}_\pm$ does not depend on the transverse coordinate, and the effective transverse area corresponds to the mode area $\pi R^2$, in complete analogy to the interaction of localized atoms with the field mode in cavity QED \cite{cavqed}.

In the stationary pulse configuration \cite{micho}, the stationary excitation is bound to the 
spin wave and
$\hmc{S}\simeq-\left(\alpha_+\hat{\psi}_++\alpha_-\hat{\psi}_-\right)$.
Solving adiabaticaly, in the limit of large optical depth $\xi$ and writing $v_g=(\alpha_+-\alpha_-)c/\eta$,
we find
\begin{subequations}\label{eq:coupspin}\begin{align}
[\partial_t+v_g\partial_z]\hmc{S} &=
i\frac{\beta}{\eta}[\hmc{S'}^\dagger\hmc{S'}]\hmc{S}
+\left[4\alpha_+\alpha_-\frac{(c\partial_z)^2}{\eta\xi}\right]
\hmc{S}
\label{eq:coupspin1}\\
\partial_t\hmc{S'} &= i\frac{\beta}{\eta}[\hmc{S}^\dagger\hmc{S}]\hmc{S'}
\label{eq:coupspin2}
\end{align}\end{subequations}
where we have ignored for now absorption and the associated noise.
\begin{figure}
\includegraphics[width=8.5cm]{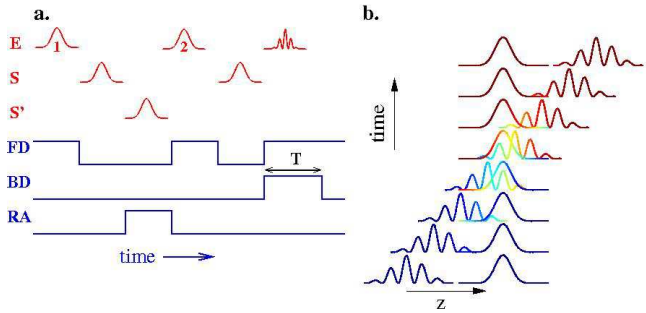}
\caption{
Nonlinear optical interaction with weak pulses as sequence of linear operations.
a. Timing diagram: forward (FD), backward (BD), and Raman or microwave (RA) intensities vs. time.
b. Illustration of Kerr interaction between slowly propagating stationary pulse $\mc{S}$ and stored excitation $\mc{S'}$ leading to phase shift (represented as changing color) of pulses.}
\end{figure} 

To solve for $\hmc{S}(z,t)$ and $\hmc{S}'(z,t)$, we first ignore pulse spreading.
Let $\hat{n}_1(z)=\hmc{S'}^\dagger\hmc{S'}$ (independent of $t$), and $\hat{n}_2(z,t)=\hmc{S}^\dagger\hmc{S}$ (which depends only on the variable $t'=t-z/v_g$). 
We find
\begin{subequations}\begin{align}
\hmc{S}(z,t) &=\exp\left[i\frac{\beta}{\eta v_g}\int\limits_{z-v_gt}^{z}dz'\;\hat{n}_1(z')\right]
\hmc{S}(z-v_gt,0)
\\
\hmc{S'}(z,t) &=\exp\left[i\frac{\beta}{\eta v_g}\int\limits_{0}^{v_gt-z}dz'\;\hat{n}_2(-z',0)\right]\hmc{S'}(z,0)
\end{align}\end{subequations}
 
When the slowly moving pulse $\hmc{S}$ has completely traversed the stored spin coherence $\hmc{S}'$,
the phase shift is $\phi_S=\frac{{\rm Re}[\beta]}{\eta v_g}L\hat{N}_{\mc{S}'}$ (where $\hat{N}_{\mc{S}'}$ is the number of excitations initially stored in $\hmc{S}'$). 
The phase shift is proportional to the interaction time, i.e. inversely proportional to the group velocity of the slowly moving pulse $\hmc{S}$.
To estimate the maximal phase shift, we note that the group velocity must be large enough that $v_gt\gtrsim l_{s'}$, where $l_{s'}$ is the length of spin coherence envelope. Also, non-adiabatic corrections due to the pulse spreading term in \eqref{eq:coupspin1} should be small, so that $\frac{(c/l_S)^2}{\eta\xi}t\lesssim 1$. Putting these two conditions together yields
\ba
\phi_S &\lesssim& d_0 \frac{\gamma}{\Delta}\frac{\sigma}{\pi R^2}\left(\frac{l_{s}^2}{L l_{s'}}\right)
\label{eq:phase}
\ea
where the resonant scattering cross-section is $\sigma=\frac{3}{4\pi}\lambda^2$.
Note that the nonlinear phase shift scales linearly with the optical depth $d_0$, in contrast to scaling with $\sqrt{d_0}$ for the case of two slowly propagating pulses \cite{lukin-ima}, in which case pulse distortion effects are also significant.

Specifically, a $300{\rm \mu m}$ long, cigar shaped cloud of cold $^{87}$Rb atoms confined in an optical dipole trap at a density of $n\sim10^{14}{\rm cm}^{-3}$ has an optical depth in excess of $d_0\sim 10^3$.  
Similar optical depth can be potentially achieved by guiding cold atom clouds of smaller density 
in a photonic crystal fiber \cite{atomguide,pbg}.
Taking the guided mode radius  $R\simeq 2{\rm \mu m}$ and accounting  for absorption losses, we choose the detuning to be $\Delta\simeq16\gamma$ and find that a  phase shift of $\phi_S\sim \pi$ is achievable due to a single stored quantum.  Under these conditions the two-photon loss probability is a few percent.


To summarize, we have shown that three-dimensional confinement of light pulses is possible by combining the technique of stationary light pulses with the transverse light guiding. This technique can be used to engineer efficient nonlinear optical interactions leading to significant phase shifts for weak optical pulses. Such interactions have interesting applications
ranging from QND measurements \cite{qnd} of few-photon pulses to quantum information processing \cite{qinfo}.

We gratefully acknowledge discussions with S. Harris and G. Kurizki.
This work was supported by DARPA, the NSF, Alfred P. Sloan and David and Lucille Packard Foundations.
 

\end{document}